\documentclass[twocolumn,showpacs,preprintnumbers,amsmath,amssymb,superscriptaddress, prl,floatfix]{revtex4-1}

\usepackage{graphicx}
\usepackage{dcolumn}
\usepackage{tabularx}
\usepackage{bm}
\usepackage{epstopdf}
\usepackage{amsmath}

\begin{document}

\title{Spinor Stochastic Resonance }

\author{H. Abbaspour}
\email[E-mail: ]{hadis.abbaspour@epfl.ch}
\affiliation{Laboratory of Quantum Optoelectronics, \'Ecole Polytechnique F\'ed\'erale de Lausanne, CH-1015, Lausanne, Switzerland}
\author{S. Trebaol}
\affiliation{UMR FOTON, CNRS, Universit\'e de Rennes 1, Enssat, F22305 Lannion, France}
\author{F. Morier-Genoud}
\affiliation{Laboratory of Quantum Optoelectronics, \'Ecole Polytechnique F\'ed\'erale de Lausanne, CH-1015, Lausanne, Switzerland}
\author{M. T. Portella-Oberli}
\affiliation{Laboratory of Quantum Optoelectronics, \'Ecole Polytechnique F\'ed\'erale de Lausanne, CH-1015, Lausanne, Switzerland}
\author{B. Deveaud}
\affiliation{Laboratory of Quantum Optoelectronics, \'Ecole Polytechnique F\'ed\'erale de Lausanne, CH-1015, Lausanne, Switzerland}
 
\date{\today}

\begin{abstract}
We report on noise-induced-spin-ordering in a collective quasipaticle system: spinor stochastic resonance. Synergetic interplay of a polarization-modulated signal and a polarization-noise allows us to switch coherently between the two metastable states of a microcavity-polariton spin bistable system. Spinor stochastic resonance is demonstrated in a zero-dimensional GaAs based microcavity. The resonance behavior of both the spin amplification and the signal-to-noise ratio are experimentally evidenced as a function of the noise strength for different amplitude modulations. They are theoretically reproduced using a spinor- Gross-Pitaevskii equation driven by a randomly polarized laser field.
\end{abstract}


\maketitle

\section{I. Introduction}
Noise is an unavoidable and random contribution in all real systems, and in particular, in their dynamics. Nevertheless, one can take advantage of such fluctuations through the counterintuitive phenomenon of \textit{stochastic resonance} : an astonishing effect that appears in nonlinear dynamical systems in which the addition of noise induces the increase of the degree of order. Since the emergence of this research field about 30 years ago \cite{Benzi1981}, stochastic resonance has been observed in a wide variety of systems \cite{Gammaitoni1998} from physics \cite{Fauve1983, Badzey2005, Abbaspour2014} to biology where the importance of such resonance has been demonstrated in living organisms \cite{Douglass1993}. Nevertheless, until know, this effect has been limited to amplitude stochastic resonance, where only the populations of the bistable states are driven. 

Polaritons are quasiparticles that arise from the strong coupling between the microcavity electromagnetic field and quantum well excitons \cite{Weisbuch1992}. Exciton-polaritons are bosonic quasiparticles and thanks to their excitonic component, they show large nonlinearities. In 2004, optical bistability has been observed in polariton systems \cite{Baas2004}. This bistability in the polariton density allowed recently the demonstration of \textit{stochastic resonance} in the polariton population inside a GaAs microcavity \cite{Abbaspour2014}. In addition, polaritons carry a spin that can be optically accessed through the polarization state of the emitted light. This allows the possible investigation of spinor polariton interactions that play a major role in the observation of spin Hall effect \cite{Leyder2007}, half-quantum vortices \cite{Lagoudakis2009}, half solitons \cite{Hivet2012}, as well as Feshbach resonance  \cite{Takemura2014}. Moreover, polariton multistability, that has been first  proposed theoretically \cite{Gippius2007} has been experimentally reported for confined polaritons \cite{Paraiso2010} : polariton strong spinor nonlinearities indeed activate the multivalued switch of a spin ensemble in a semiconductor microcavity. This has shown great potential to investigate coherent spin dynamics  \cite{Amo2010, Cerna2013}. 

During the last decade, a number of studies have focused on optical devices based on the non-linearities of polaritons \cite{Amo2010, Cerna2013, Tsintzos2008, Ballarini2013, Grosso2014}. In this framework, noise inherently impacts the overall dynamics of systems and should be studied thoroughly. Recently, intensity noise measurements in polariton gas have been reported both on theoretical \cite{Glazov2013} and experimental aspects \cite{Poltavtsev2014, Boulier2014}, with particularly the demonstration of the stochastic resonance in the polariton population \cite{Abbaspour2014}. More than just an intriguing phenomenon, stochastic resonance might be used as a tool to increase the sensitivity of nonlinear devices and to extract signal information from a noisy environment  \cite{Murali2009}. 

Spintronics and, more recently, spinoptronics  \cite{Amo2010, Liew2008} are research fields demonstrating innovative devices that take advantage of the quantum properties of the spin. Some works on stochastic resonance, based on spin control, have been reported in magnetic materials  \cite{Cheng2010, Finocchio2011, Gourier1998}. In particular, stochastic resonance has been demonstrated in a nanoscale spin-valve driven by spin-polarized current  \cite{Cheng2010}. Nevertheless until now the demonstration of a spin ordering induced by noise is still missing. 

Here we demonstrate a new effect in stochastic resonance by showing that polarization noise may fully switch the spin state of the polariton population. The mechanism involves a special polariton bistable behavior deeply linked with the existence of a biexciton resonance. We take here advantage of the unique spin properties of exciton-polaritons to achieve the noise-induced spin-ordering in collective exciton-polariton excitations within a 0-dimensional (0D) semiconductor microcavity. We name this phenomenon ``\textit{spinor stochastic resonance}''. This allows us to unveil an original field of stochastic resonance based on the ordering of the spin of a collective ensemble of particles by spin noise. The effect is evidenced by the spin amplification of the noisy modulated input signal through the enhancement of the degree of polarization up to a fully circular polarized light. Concomitant with its intrinsic interest, spinor stochastic resonance might get into broad investigations on the effect of fluctuations on spinoptronic devices and allow to propose schemes taking advantage of inherent noise contributions.

The paper is organized as follows: In Section II, we describe the principle of the stochastic resonance. Section III reports on the experimental method and the demonstration of the spinor stochastic resonance. Section IV is dedicated to the theoretical model based on spinor Gross-Pitaevskii equation driven by stochastic excitation. We give our conclusions in Section V.
\section{II. The principle of the stochastic resonance}
The basic principle of the stochastic resonance is the following. Let’s consider a Brownian particle in a double well potential, which is initially located in one of the two wells. Notice that the double well potential corresponds to a bistable system. Thermal noise will induce hopping of the particle between the two minima of the potential  \cite{Gammaitoni1998}. The switching occurrence, called the Kramers rate  \cite{Kramers1940}, directly depends on the noise level. The larger the noise amplitude, the shorter will be the Kramers time of the particle in each well. We consider now a periodic force applied on the double well potential with a small enough modulation amplitude to avoid any deterministic hopping of the particle between the two wells. For some optimal noise intensity, the Kramers rate matches roughly the modulation frequency of the external force, and consequently, synchronization takes place between the periodic signal and the noise therefore inducing quasi deterministic jumps. At this noise value, we reach \textit{stochastic resonance}. 
\section{III. Experimental method}
We consider a spin-trigger regime  \cite{Paraiso2010, Cerna2013} that simply derives from a double well potential featuring two spin minima: spin-up ($\uparrow$) and spin-down ($\downarrow$). We reach the polariton spin-trigger regime thanks to spinor polariton nonlinearity and multistability in confined polaritons in 3 $\mu$m diameter mesa in a GaAs planar microcavity. These 0D polaritons \cite{ElDaif2006, Nardin2009} actually display a discrete energy spectrum  and present key parameter for achieving polarization multistability and therefore spinor stochastic resonance: a narrow linewidth of the polariton ground state ($\gamma =100 \mu$eV), which is isolated from internal fluctuations. The sample is cooled down to 4K and we carry out the experiments at exciton-cavity detuning $\delta=0.1$ meV. We excite the sample using cw single mode Ti:Sapphire laser with 20 $\mu$m diameter. The laser energy is blue detuned from the polariton ground state ($\Delta=0.8$ meV). Using an electro-optic modulator we can apply an external modulation or Gaussian noise on the polarization of the laser beam. The transmitted signal is projected into the circular polarization basis using a quarter-wave plate, that converts spin-up and spin-down populations into cross-linear polarization. We separate the two linear polarizations using a polarizing beam splitter and record them simultaneously with two 20 MHz bandwidth photodiodes connected to a 60 MHz bandwidth oscilloscope. We present the results based on excitation ($\rho_{in}$) and emission ($\rho_{out}$) circular polarization degree defined as:
\begin{equation}
\rho_{in, out}=\frac{I_{\sigma+}-I_{\sigma-}}{I_{\sigma+}+I_{\sigma-}} 
\end{equation}
\begin{figure}[tb] 
\centering 
\includegraphics[width=1\columnwidth]{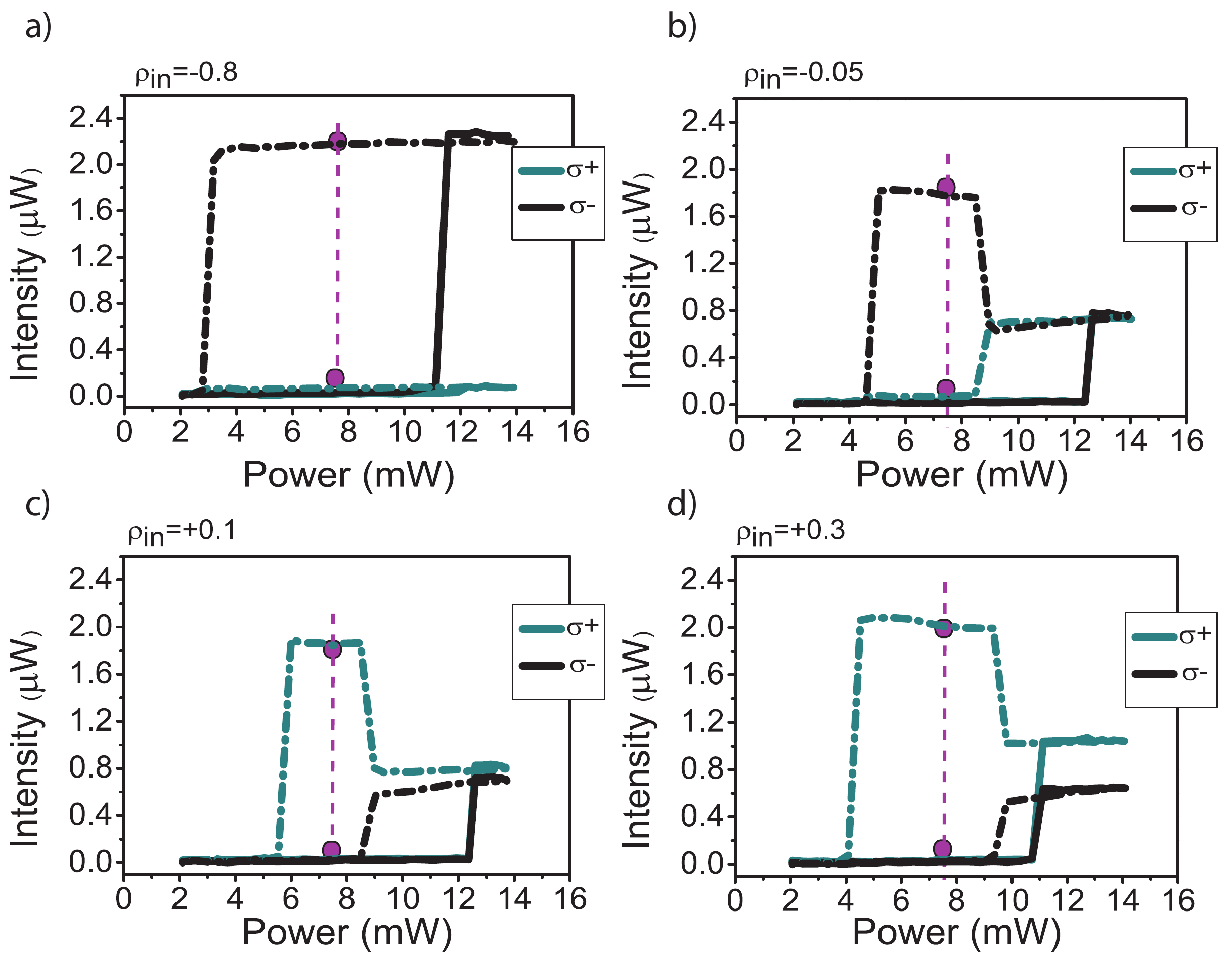} 
\caption[Spinor bistability measurement]{ (Color online) \textbf{ Spinor bistability measurement.} The excitation power is scanned from 2 mW to 14 mW forward and backward for four given polarization degrees $\rho_{in}$=-0.8, -0.05, +0.1 and +0.3. \textbf{a} At large circular polarization degree ($\rho_{in}$=-0.8), the system shows a conventional polariton bistability. \textbf{b-d} The possibility of biexciton creation gives rise to a nonliner loss mechanism in the polariton system. This results in the middle stable branch around 10 mW. When the minority spin population falls down to its lower state, the biexciton formation mechanism decreases and the majority spin population remains in resonance with the laser. As a consequence, the majority polariton intensity jumps up. For the following experiments, the laser power is fixed at 7.8 mW (dashed line).  Spinor bistabilities are obtained for exciton-cavity detuning ($\delta=0.1$ meV) and a polariton-laser detuning of ($\Delta=0.8$ meV).}
\end{figure}
\begin{figure}[b] 
\centering 
\includegraphics[width=1\columnwidth]{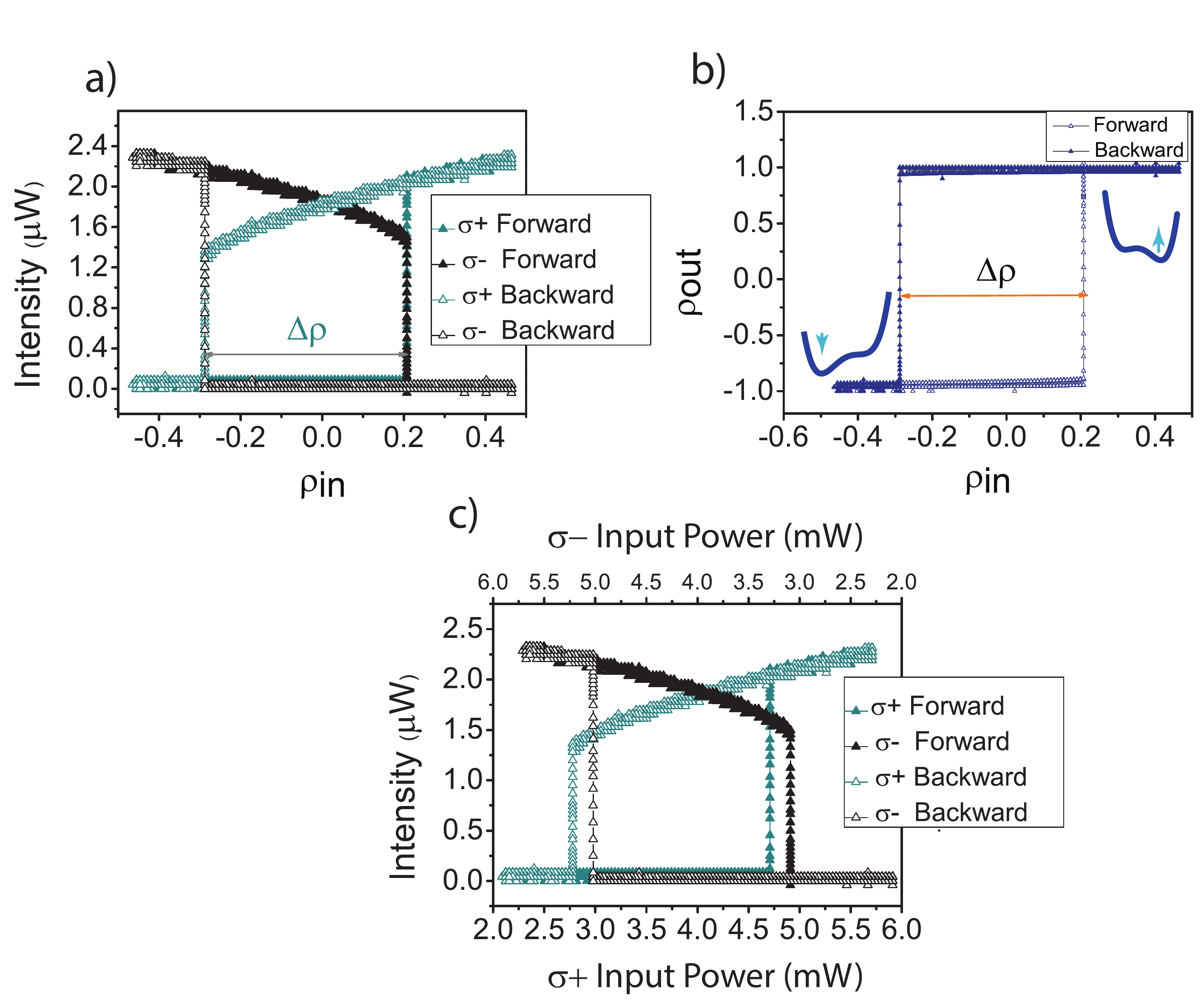} 
\caption[Spin-trigger regime]{(Color online) \textbf{Spin-trigger regime.} For the fixed excitation power of 7.8 mW (see Fig. 1), the laser polarization degree is scanned between -0.45 and 0.45 forward and backward. Two bistabilities are measured using the usual X-Y method \cite{Smith1977} that directly links input and outut signals. \textbf{a} Spin-up and spin-down polariton populations as a function of the  excitation polarization degree show two overlapping bistabilities. \textbf{b} The polariton polarization degree versus input polarization shows the hysteresis loop called spin-trigger regime ($\Delta$$\rho$=0.5).  The polariton population switches from spin-down ($\rho_{out}$=-1) to spin-up ($\rho_{out}$=+1) states with a hysteresis in input polarization ($\rho_{in}$). \textbf{c} Both polariton spin populations versus effective laser power for each spin-up (lower axis) and spin-down (upper axis) polaritons. The curves evidence two polariton bistability loops with the same width $\Delta$B=1.93 mW.}
\end{figure}
\subsection{A. Spin-Trigger}
In order to investigate spinor stochastic resonance, we obtain the spin-trigger regime as follows. First, scanning the input laser intensity for a fixed input polarization we measure the transmitted intensities of  both spin up ($I_{\sigma+}$) and spin down  ($I_{\sigma-}$) polaritons. This step is reproduced for different input polarizations ($\rho_{in}$) between +1 and -1 revealing an ensemble of polariton multistability behaviors (Fig. 1). For a laser polarization close to the fully left circular state ($\rho_{in}$=-0.8), we observe a usual polariton bistability for the majority polariton population ($I_{\sigma-}$) (Fig. 1 (a)). Figure 1 (b) to (d) shows the bistability loop of both polariton spin populations for a laser polarization close to linear. We observe that the two upper thresholds coincide, while the lower thresholds are decoupled. The presence of a biexciton reservoir, closely resonant to the input laser energy, is primordial to obtain this type of spinor bistability \cite{Wouters2013}. This condition is satisfied with the exciton-cavity and laser-ground state detuning used in the experiment. Indeed, biexciton creation, resulting from the combination of one spin-up and one spin-down polariton, is resonantly enhanced at the reservoir energy. In fact, when decreasing the excitation power, the minority spin population falls down to the lower branch. Accordingly, the biexciton creation mechanism decreases, and the majority spin population increases. Then, for further reduction of laser power, the majority spin population falls down also to the lower branch. 

Second, we identify the laser intensity for which polaritons are either on the spin-up or spin-down state of the hysteresis. Then for this fixed laser power, which is here 7.8 mW, we tune the polarization degree ($\rho_{in}$) from circular-left (-1) to circular-right (+1) favoring respectively the creation of spin-up or spin-down polaritons. 

In Figure 2(a), we show the evolution of the two polariton spin populations obtained when scanning the laser polarization degree ($\rho_{in}$) between -0.45 and +0.45 forwards and backwards. Using equation (1) we have calculated polariton polarization degree ($\rho_{out}$). The detected polarization state displays a clear hysteresis behaviour, directly imaging the bistability of the spin state of polaritons (Fig. 2(b)). For $\rho_{in}$=-0.45, polaritons are in the spin-down state. Upon changing the polarization of the excitation beam towards the right-circular direction, at $\rho_{in}$=0.21, polaritons sharply jump to the spin-up state. Then by sweeping the ellipticity in the backward direction, a second threshold occurs at $\rho_{in}$=-0.29 and polaritons turn back to the spin-down state. Under such conditions we reach the polariton spin-trigger regime, in which it is possible to switch between the two well-defined  spin orientations of the polariton population : spin-up $\uparrow$ and spin-down $\downarrow$ with a large hysteresis width  $\Delta$$\rho$=0.5. In fact, the polariton nonlinear loss, due to the biexciton formation, only allows the dominant spin population to be transmitted, while the minority population is totally absorbed. This is imprinted onto the emitted light, which flips between fully circularly polarized states.
\begin{figure}[b]
\includegraphics[width=1\columnwidth]{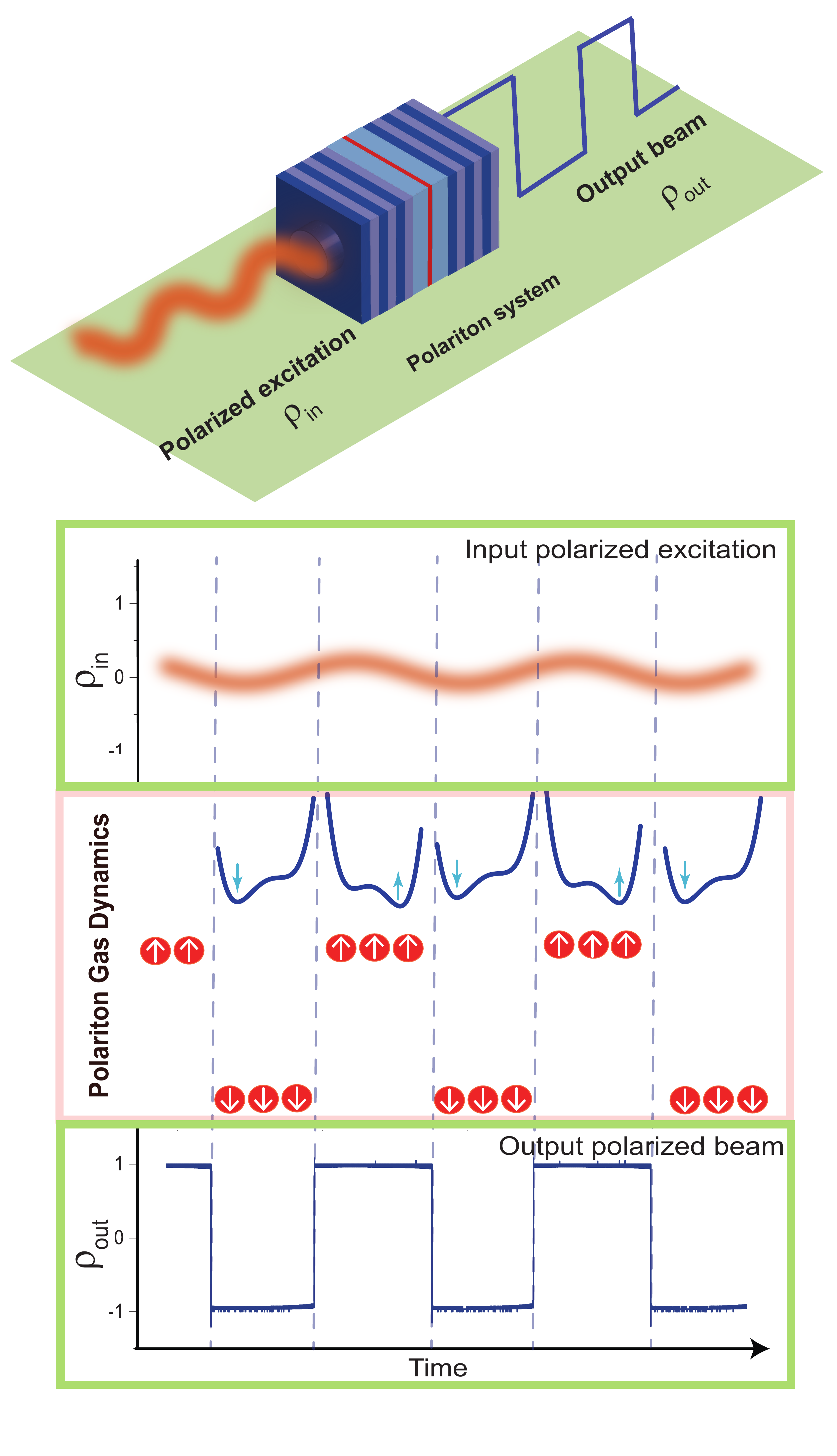}
\caption{(Color online) Principle of spinor stochastic resonance. Spin amplification: A noisy polarization modulated signal around linear polarization state controls the polariton population between two well-defined spin states $\pm 1$. The spin-trigger regime can be simply modeled as a double well potential in two spin states $\uparrow$ and $\downarrow$. Altering the input polarization $\rho_{in}$, favors one of these wells compared to the other one. This induces polarization ordering of polariton population in time with a defined spin either up or down. As a result, the emitted light has full circular polarization which alternates between $\sigma$+ and $\sigma-$. }
\end{figure}
\begin{figure*}[tb] 
\centering 
\includegraphics[width=\textwidth,height=16cm]{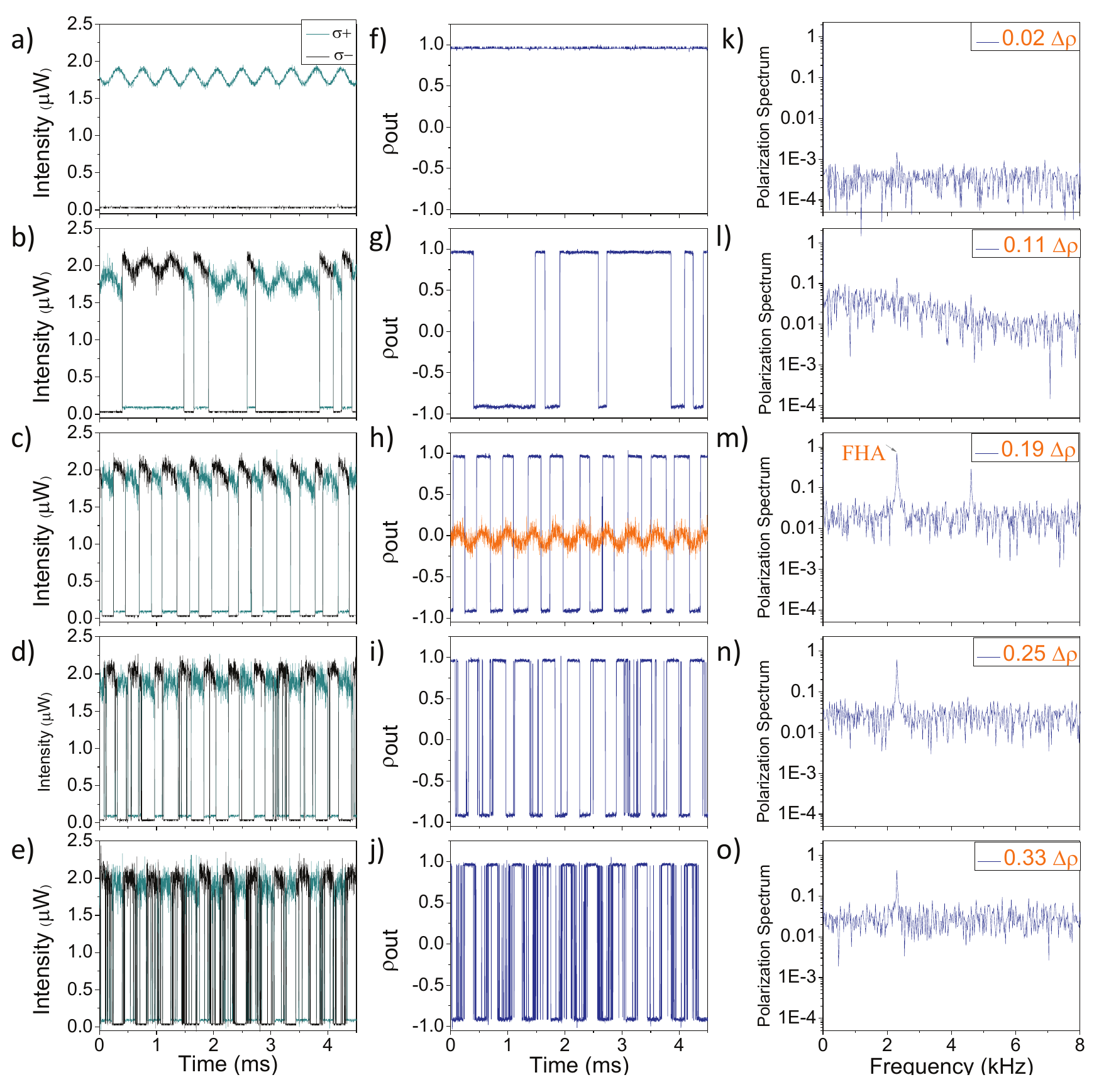} 
\caption[Demonstration of spinor stochastic resonance]{(Color online) \textbf{Demonstration of spinor stochastic resonance.} \textbf{a-e} Spin-up (green) and spin-down (black) polariton population in time domain for different polarization fluctuations  ($D_{\sigma}$) of : 39 $\mu$W (0.02 $\Delta$B), 212 $\mu$W (0.11 $\Delta$B), 367 $\mu$W (0.19 $\Delta$B), 482 $\mu$W (0.25 $\Delta$B), 637 $\mu$W (0.33 $\Delta$B). \textbf{f-j} Polariton spinor state in time domain (corresponding to (a-e)) for different polarization noise ($D_{\rho_{in}}$) : 0.02 $\Delta$$\rho$, 0.11 $\Delta$$\rho$, 0.19 $\Delta$$\rho$, 0.25 $\Delta$$\rho$, 0.33 $\Delta$$\rho$. \textbf{k-o} Frequency spectra corresponding to (f-j). The polarization amplitude of the input signal is $A_0=0.085 (0.17 \Delta$$\rho$) and the frequency of modulation $\nu_0=2.33$ kHz. At an optimal noise (0.19 $\Delta$$\rho$) periodic switching between two spin states occurs (spinor stochastic resonance). In (h) the input periodic polarization signal is superimposed to the output signal. In (m) first-harmonic amplitude FHA is the amplitude of the transmitted signal at the input modulation frequency.}
\end{figure*}
The polarization degree of the excitation laser determines the effective power which drives each spin population ($F_{\pm}$), even though their addition would be constant at fixed laser power (7.8 mW). Figure 2(c) shows the intensity of spin up ($I_{\sigma+}$) and spin down ($I_{\sigma-}$) polaritons as a function of the driving laser intensity. We start with $\rho_{in}$=-0.45, which corresponds to ($F_{+}$=2.15 mW) spin up and ($F_{-}$=5.65 mW) spin down excitation laser. By increasing the $\sigma+$ contribution of the laser, for a certain amount of laser polarization ($F_{+}$=3.09 mW), spin down polaritons fall back to the lower state and spin up polaritons jump in resonance with the laser. Then by changing the laser polarization in the backward direction, we observe the second threshold at ($F_{-}$=2.77 mW). Finally we observe two polariton bistabilities for two different spin populations with the same width of $\Delta$B=1.93 mW. The small dissymmetry between the two polariton bistabilities is linked with the splitting of the confined polariton states in the mesa structure.
\subsection{B. Spinor Stochastic Resonance}
In order to study spinor stochastic resonance, we fix the polarization state of the input beam at $\rho_{in}$=-0.04 to drive the system in the middle of the spin-trigger hysteresis loop (Fig. 2(b)). Then, we sweep the polarization of the input signal with a small sinusoidal modulation amplitude $A_0<0.5 \Delta$$\rho$, which does not permit polaritons to jump to the other bistable spin state. Adding a proper amount of noise in polarization to the sinusoidal signal allows polaritons to overcome non-linear thresholds and to display well defined spin states in transmission. This is the principle of the \textit{spinor stochastic resonance}, which is illustrated schematically in Fig. 3. 

Practically, we imprint a sinusoidal modulation amplitude $A_0$ at frequency $\nu_0=2.33$ kHz and a 500 kHz Gaussian noise in polarization on the input laser beam using an electro-optic modulator. It is worth mentioning that since the noise bandwidth is two orders of magnitude broader than the modulation frequency, the applied noise can be considered as a white noise. To investigate the role of the strength of polarization noise on the output signal, the amplitude of the white noise is adjusted from 0 to 0.6 $\Delta$$\rho$. Practically, we measure the polarization noise ($D_{\rho_{in}}$), which is imprinted on the  laser, through the intensity noise ($D_{\sigma}$) of both spin populations. Since the two intensity fluctuations are anticorrelated, the effective noise is equivalent for both spin-up and spin-down populations. Then, using equation (1) we calculate the spin noise in the system. We normalize the intensity noise for spin-up and spin-down laser intensity by the bistability width ($\Delta$B) in Figure 2(c). The spin noise values are also normalized by the spin-trigger width ($\Delta$$\rho$) (Fig. 2(b)). While increasing the polarization noise standard deviation ($D_{\rho_{in}}$), we record simultaneously the polariton emission intensities in the circular basis $I_{\sigma+}$ and $I_{\sigma-}$ as a function of time, and then compute circular polarization degree of polariton population $\rho_{out}$ using Eq. (1). Figure 4 displays the results recorded for an amplitude of the modulated signal $A_0=0.17 \Delta$$\rho$. The noise $D_{\rho_{in}}$ ($D_{\sigma}$) is changed from 0.02 to 0.33 $\Delta$$\rho$ ($\Delta B$).

 We show in Fig. 4 (a)-(e) the effect of the increased noise on the dynamics of the two polariton spin populations, and the corresponding time dependence of the polarization state (Fig. 4 (f)-(j)). The corresponding frequency spectra are shown in Fig. 4 (k)-(o). The system is initialized in the upper state $\uparrow$ of the spin trigger. For the minimum noise value ($D_{\rho_{in}}$=0.02 $\Delta$$\rho$, $D_{\sigma}$=0.02 $\Delta$B), the periodic force is not large enough to overcome the non-linear thresholds and the system stays in the spin up state (Fig. 4 (a), (f)). Moreover, the small slope of this upper branch of the spin-trigger hysteresis (Fig. 2 (b)) prevents the transmission of the input modulation. By increasing the noise amplitude to $D_{\rho_{in}}$=0.11 $\Delta$$\rho$, $D_{\sigma}$=0.11 $\Delta$B, the addition  of  external fluctuations to the  periodic signal  starts to induce erratic jumps between spin-up and spin-down states (Fig. 4 (b), (g)). For the optimal noise condition  ($D_{\rho_{in}}$=0.19 $\Delta$$\rho$, $D_{\sigma}$=0.19 $\Delta$B), the interplay between external forces switches the coherent polariton population between the two spin-up and spin-down states with the same frequency as the external modulation  $\nu_0$. Therefore, fluctuations authorize controlling polariton spin population inside the microcavity. To emphasize the frequency locking between input and output signal, which is a characteristic of stochastic resonance, we superimpose the sinusoidal input on Figure 4 (h). Here we observe experimentally the spinor stochastic resonance behavior described on Figure 3. A noisy modulated polarized input signal coherently controls a spinor polariton ensemble between well-defined spin states. For still larger noise amplitudes, the synchronization progressively disappears (Fig. 4 (d)-(e), (i)-(j)) and the periodic jumps start to be hidden in the noise.\\

The amplitude of the transmitted signal at the input modulation frequency $\nu_0$ is the key parameter to evidence stochastic resonance \cite{Badzey2005, Abbaspour2014}. We therefore Fourier transform polariton polarization time streams recorded for a 50 ms period to obtain polarization spectra with a spectral resolution of 21 Hz. In the third column of Figure 4 we display the polarization spectra corresponding to the time stream presented in the second column. From each spectrum, we extract the first harmonic amplitude (FHA), which reaches its maximum value at $D_{\rho_{in}}$=0.19 $\Delta$$\rho$ (Fig. 4 (m)). Note that for small noise amplitude, since no modulation is transmitted (Fig. 4 (f)), no clear peak appears at $\nu_0$ (Fig. 4 (k)). 
\subsection{Spin Amplification}
To reveal the stochastic resonance, we study, for different amplitudes of the modulation signal, the polariton spin amplification 
\begin{equation}
\eta=\frac{FHA}{FHA_{in}} 
\end{equation}
as a function of the spin noise $D_{\rho_{in}}$. FHA and $FHA_{in}$ are the first harmonic amplitude in the polarization spectrum of the output signal and the corresponding minimum noise input signal, respectively. In Figure 5(a) we plot the spin amplification as a function of noise for input signal modulation amplitudes of 0.17  $\Delta$$\rho$, 0.26  $\Delta$$\rho$ and 0.39  $\Delta$$\rho$. Their extracted $FHA_{in}$ are 0.07, 0.1 and 0.18 degree of polarization respectively. \\
\begin{figure}[tb] 
\includegraphics[scale=0.35]{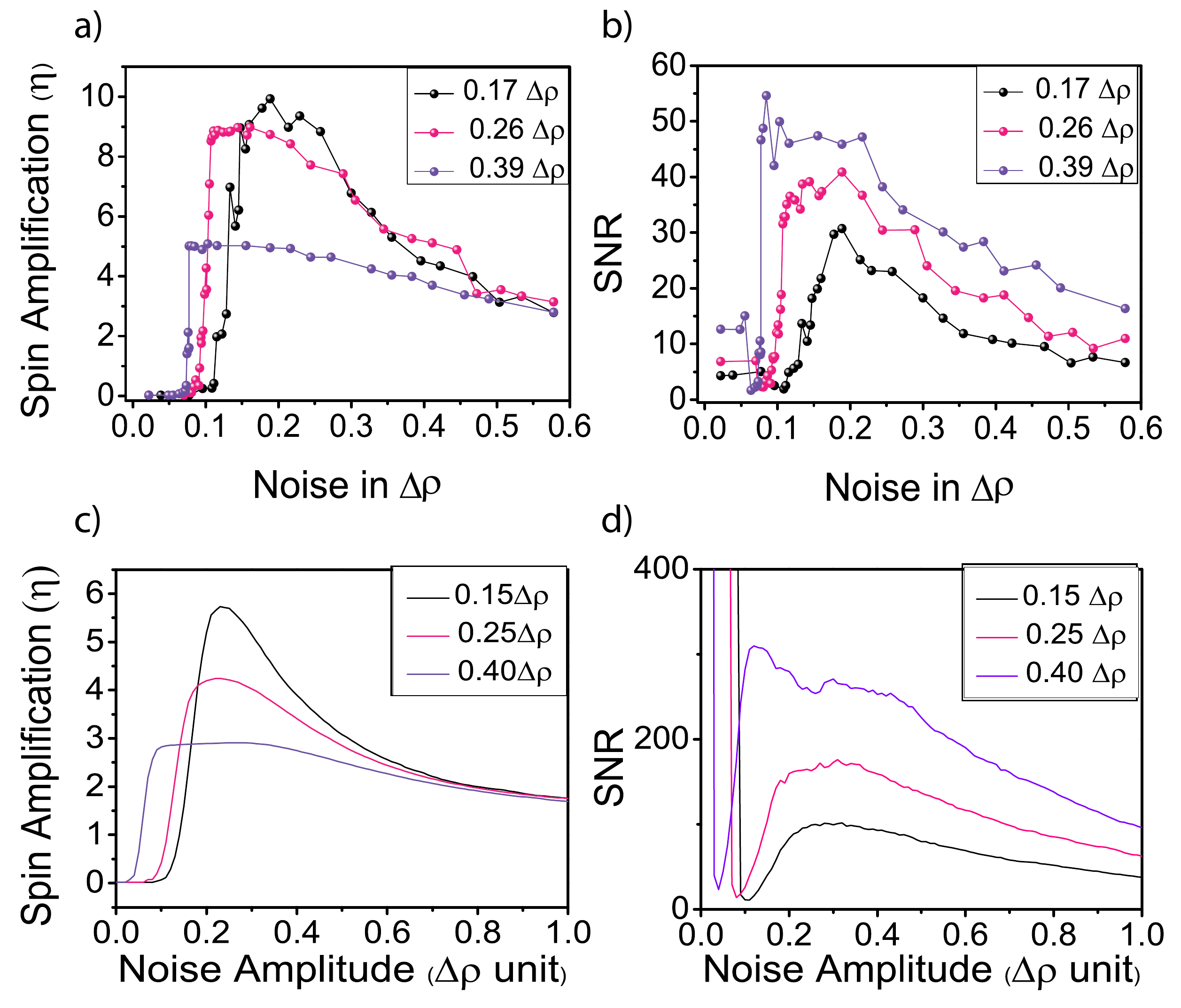}
\caption{(Color online) \textbf{a} Experimental and \textbf{c} numerical spin amplification ($\eta$), \textbf{b} Experimental and \textbf{d} numerical Signal-to-noise ratio (SNR) for three different modulation amplitudes $A_0$. $\Delta$$\rho$ is the polariton spin bistability width. The resonant behavior is evident for both quantifiers. }
\end{figure}
For the same modulation amplitude as in Figure 4 ($A_0=0.17 \Delta$$\rho$), we observe a large spin amplification, reaching the stochastic resonance condition at $D_{\rho_{in}}$=0.19 $\Delta$$\rho$. This behavior clearly evidences the essential role played by the spin noise on the transmission of the carried input signal at the frequency $\nu_0$. Eventually, synergic interplay between noise and modulation in spin domain assists the system to amplify the input polarization through the microcavity. This specific amplification profile is expected for a bistable potential \cite{Gammaitoni1998}. 

We now focus on the influence of the applied modulation amplitude on the spinor stochastic resonance. Increasing the modulation amplitude to  0.26  $\Delta$$\rho$ and 0.39  $\Delta$$\rho$, the amplification decreases accordingly and the resonance shape is less pronounced. The larger the polarization amplitude, the easier it is to achieve deterministic jumps of polariton population between the two spin states. Consequently the amount of noise allowing the observation of stochastic resonance decreases and the system tends to show a flat response for different noise amplitudes.
\subsection{Signal-to-Noise Ratio}
To evaluate the role of the spin noise, we study for the same modulated-signal-polarization amplitudes, the signal-to-noise ratio (SNR) defined as
\begin{equation}
SNR=\frac{FHA}{N(\nu,D)} 
\end{equation}
where  $N(\nu,D)$ is the noise background averaged between 2.42 kHz and 3.83 kHz. The SNRs are presented in Figure 5(b). Similar to the polariton spin amplification, we notice the expected stochastic resonance behavior as a function of noise, being more peaked for minimum amplitude $A_0=0.17 \Delta$$\rho$ than for  0.26  $\Delta$$\rho$ and 0.39  $\Delta$$\rho$. In fact, for larger $A_0$ deterministic jumps of polaritons between the two spin states are favored. This implies an increase of the SNR accompanied by a decreasing of the $D_{\rho_{in}}$ value needed to reach the stochastic resonance. Finally, comparing Figure 5(a) and (b), one can notice that the spin amplification at the stochastic resonance always goes together with a recovering of the SNR. 
\section{IV. theoretical model}
The dynamical description of two spin mode polariton wave functions ($\Psi_{\uparrow, \downarrow}$) can be obtained in the circular basis ($\sigma+$, $\sigma-$) by using spinor Gross-Pitaevskii equation \cite{Wouters2013} driven by a randomly polarized input :
\begin{equation}
\begin{split}
i \frac{d\Psi_{\uparrow, \downarrow}}{dt}&=[-\Delta-\frac{i}{2}(\gamma+\beta|\Psi_{\downarrow, \uparrow}|^2)+ \alpha_1|\Psi_{\uparrow, \downarrow}|^2+\\
&\alpha_2|\Psi_{\downarrow, \uparrow}|^2]\Psi_{\uparrow, \downarrow}+\frac{\epsilon_{lin}}{2}\Psi_{\downarrow,\uparrow}+F_{\uparrow, \downarrow}
\label{GPE}
\end{split}
\end{equation}
where $\beta$, $\epsilon_{lin}$ and $\gamma$ represent the biexciton nonlinear loss, linear polarization splitting and polariton linewidth respectively. We consider anisotropic spin interactions between co-polarized ($\alpha_1$) and cross-polarized ($\alpha_2$) polaritons. $\Delta$ is the energy detuning between the polariton ground state and the laser. Driving fields $F_{\uparrow, \downarrow}=\sqrt{I_{\uparrow, \downarrow}}$ for the two polariton polarizations, are defined as
\begin{equation}
I_{\uparrow, \downarrow}(t')=\frac{I_0}{2}\times|\rho_{in}\pm 1+A_0  \cos(2\pi\nu_0t^{\prime}+\phi)+D(t^{\prime})|
\end{equation}
where D, the polarization noise amplitude, follows a normal distribution with standard deviation $D_{\rho_{in}}$ and $\mid\rho_{in}\mid<0.25$ is the DC polarization component of the laser excitation.  $I_0$  is the fixed laser intensity expressed as  $I_0=I_{\sigma+}+I_{\sigma-}$. As the modulation frequency is orders of magnitude slower than the intrinsic polariton dynamics (few GHz), we can apply the adiabatic approximation \cite{Abbaspour2014} and consider $F_{\uparrow, \downarrow}$ as constant driving terms in equation (4). $F_{\uparrow, \downarrow} (t^{\prime})$ is an input time sequence of 1 second with time steps of 2 $\mu$s, corresponding to the noise correlation time. Here the polarization noise can be approximated as a white noise since the modulation frequency is only 2.33 kHz.

The results of our numerical simulations for the spin amplification and the SNR are presented in Figure 5(c) and (d) respectively. All the experimental features described above
are qualitatively reproduced by our model. Notice that Figure 5(b) shows smaller amount of SNR compared to numerical results. This can be due to experimental background noise as well as internal noise in the system, which are not included in the model. As expected, the computed SNR diverges for vanishing polarization fluctuations, while residual noise of the laser prevents experimental SNR to display such large values.
\section{V. conclusion}
Using the polariton spin-trigger regime, we demonstrated spinor stochastic resonance. Addition of polarization noise on a polarization modulated signal around a linear polarization state induces spin ordering of the polariton population, in which the spin of the collective polariton excitation alternates  periodically between spin-up and spin-down at the frequency of the modulation signal. As a result, the emitted light alternates between $\sigma$+ and $\sigma-$  with full circular polarization. We demonstrated experimentally and reproduced theoretically the resonance for the amplification and for the signal-to-noise ratio as a function of noise strength for different amplitude modulation. Spinor stochastic resonance might be a new tool for improving spintronic devices.
\\\\
The present work has been supported by the Swiss National Science Foundation under Project No. N135003, the Quantum Photonics National Center of Competence in Research under Project No. N115509, and by the European Research Council under Project POLARITONICS Contract No. N219120. The POLATOM Network is also acknowledged.


\begin{thebibliography}{10}%
\makeatletter
\providecommand \@ifxundefined [1]{%
 \ifx #1\undefined \expandafter \@firstoftwo
 \else \expandafter \@secondoftwo
\fi
}%
\providecommand \@ifnum [1]{%
 \ifnum #1\expandafter \@firstoftwo
 \else \expandafter \@secondoftwo
\fi
}%
\providecommand \enquote [1]{``#1''}%
\providecommand \bibnamefont  [1]{#1}%
\providecommand \bibfnamefont [1]{#1}%
\providecommand \citenamefont [1]{#1}%
\providecommand\href[0]{\@sanitize\@href}%
\providecommand\@href[1]{\endgroup\@@startlink{#1}\endgroup\@@href}%
\providecommand\@@href[1]{#1\@@endlink}%
\providecommand \@sanitize [0]{\begingroup\catcode`\&12\catcode`\#12\relax}%
\@ifxundefined \pdfoutput {\@firstoftwo}{%
 \@ifnum{\z@=\pdfoutput}{\@firstoftwo}{\@secondoftwo}%
}{%
 \providecommand\@@startlink[1]{\leavevmode}%
 \providecommand\@@endlink[0]{}%
}{%
 \providecommand\@@startlink[1]{%
  \leavevmode
  \pdfstartlink
   attr{/Border[0 0 1 ]/H/I/C[0 1 1]}%
   user{/Subtype/Link/A<</Type/Action/S/URI/URI(#1)>>}%
  \relax
 }%
 \providecommand\@@endlink[0]{\pdfendlink}%
}%
\providecommand \url  [0]{\begingroup\@sanitize \@url }%
\providecommand \@url [1]{\endgroup\@href {#1}{\urlprefix}}%
\providecommand \urlprefix [0]{URL }%
\providecommand \Eprint[0]{\href }%
\@ifxundefined \urlstyle {%
  \providecommand \doi [1]{doi:\discretionary{}{}{}#1}%
}{%
  \providecommand \doi [0]{doi:\discretionary{}{}{}\begingroup
  \urlstyle{rm}\Url }%
}%
\providecommand \doibase [0]{http://dx.doi.org/}%
\providecommand \Doi[1]{\href{\doibase#1}}%
\providecommand \bibAnnote [3]{%
  \BibitemShut{#1}%
  \begin{quotation}\noindent
    \textsc{Key:}\ #2\\\textsc{Annotation:}\ #3%
  \end{quotation}%
}%
\providecommand \bibAnnoteFile [2]{%
  \IfFileExists{#2}{\bibAnnote {#1} {#2} {\input{#2}}}{}%
}%
\providecommand \typeout [0]{\immediate \write \m@ne }%
\providecommand \selectlanguage [0]{\@gobble}%
\providecommand \bibinfo [0]{\@secondoftwo}%
\providecommand \bibfield [0]{\@secondoftwo}%
\providecommand \translation [1]{[#1]}%
\providecommand \BibitemOpen[0]{}%
\providecommand \bibitemStop [0]{}%
\providecommand \bibitemNoStop [0]{.\EOS\space}%
\providecommand \EOS [0]{\spacefactor3000\relax}%
\providecommand \BibitemShut [1]{\csname bibitem#1\endcsname}%
\bibitem{Benzi1981}%
  \BibitemOpen
  \bibfield{author}{%
  \bibinfo {author} {\bibfnamefont{R.}~\bibnamefont{Benzi}}, \bibinfo {author}
  {\bibfnamefont{A.}~\bibnamefont{Sutera}},\ and\ \bibinfo {author}
  {\bibfnamefont{A.}~\bibnamefont{Vulpiani}},\ }%
  \bibfield{journal}{%
  \bibinfo {journal} {Journal of Physics A}\ }%
  \textbf{\bibinfo {volume} {14}} (\bibinfo {year} {1981})%
  \bibAnnoteFile{NoStop}{Benzi1981}%
\bibitem{Gammaitoni1998}%
  \BibitemOpen
  \bibfield{author}{%
  \bibinfo {author} {\bibfnamefont{L.}~\bibnamefont{Gammaitoni}}, \bibinfo
  {author} {\bibfnamefont{P.}~\bibnamefont{H{\"a}nggi}}, \bibinfo {author}
  {\bibfnamefont{P.}~\bibnamefont{Jung}},\ and\ \bibinfo {author}
  {\bibfnamefont{F.}~\bibnamefont{Marchesoni}},\ }%
  \bibfield{journal}{%
  \bibinfo {journal} {Rev. Mod. Phys.}\ }%
  \textbf{\bibinfo {volume} {70}},\ \bibinfo {pages} {223} (\bibinfo {year}
  {1998})%
  \bibAnnoteFile{NoStop}{Gammaitoni1998}%
\bibitem{Fauve1983}%
  \BibitemOpen
  \bibfield{author}{%
  \bibinfo {author} {\bibfnamefont{S.}~\bibnamefont{Fauve}}\ and\ \bibinfo
  {author} {\bibfnamefont{F.}~\bibnamefont{Heslot}},\ }%
  \bibfield{journal}{%
  \bibinfo {journal} {Phys. Lett. A}\ }%
  \textbf{\bibinfo {volume} {97}},\ \bibinfo {pages} {5} (\bibinfo {year}
  {1983})%
  \bibAnnoteFile{NoStop}{Fauve1983}%
\bibitem{Badzey2005}%
  \BibitemOpen
  \bibfield{author}{%
  \bibinfo {author} {\bibfnamefont{R.~L.}\ \bibnamefont{Badzey}}\ and\ \bibinfo
  {author} {\bibfnamefont{P.}~\bibnamefont{Mohanty}},\ }%
  \bibfield{journal}{%
  \bibinfo {journal} {Nature}\ }%
  \textbf{\bibinfo {volume} {437}},\ \bibinfo {pages} {995} (\bibinfo {year}
  {2005})%
  \bibAnnoteFile{NoStop}{Badzey2005}%
\bibitem{Abbaspour2014}%
  \BibitemOpen
  \bibfield{author}{%
  \bibinfo {author} {\bibfnamefont{H.}~\bibnamefont{Abbaspour}}, \bibinfo
  {author} {\bibfnamefont{S.}~\bibnamefont{Trebaol}}, \bibinfo {author}
  {\bibfnamefont{F.}~\bibnamefont{Morier-Genoud}}, \bibinfo {author}
  {\bibfnamefont{M.~T.}\ \bibnamefont{Portella-Oberli}},\ and\ \bibinfo
  {author} {\bibfnamefont{B.}~\bibnamefont{Deveaud}},\ }%
  \bibfield{journal}{%
  \Doi{10.1103/PhysRevLett.113.057401}{\bibinfo {journal} {Phys. Rev. Lett.}}\
  }%
  \textbf{\bibinfo {volume} {113}},\ \bibinfo {pages} {057401} (\bibinfo
  {month} {Jul}\ \bibinfo {year} {2014})%
  \bibAnnoteFile{NoStop}{Abbaspour2014}%
\bibitem{Douglass1993}%
  \BibitemOpen
  \bibfield{author}{%
  \bibinfo {author} {\bibfnamefont{J.~K.}\ \bibnamefont{Douglass}}, \bibinfo
  {author} {\bibfnamefont{L.}~\bibnamefont{Wilkens}}, \bibinfo {author}
  {\bibfnamefont{E.}~\bibnamefont{Pantazelou}},\ and\ \bibinfo {author}
  {\bibfnamefont{F.}~\bibnamefont{Moss}},\ }%
  \bibfield{journal}{%
  \bibinfo {journal} {Nature}\ }%
  \textbf{\bibinfo {volume} {365}},\ \bibinfo {pages} {337} (\bibinfo {year}
  {1993})%
  \bibAnnoteFile{NoStop}{Douglass1993}%
\bibitem{Weisbuch1992}%
  \BibitemOpen
  \bibfield{author}{%
  \bibinfo {author} {\bibfnamefont{C.}~\bibnamefont{Weisbuch}}, \bibinfo
  {author} {\bibfnamefont{M.}~\bibnamefont{Nishioka}}, \bibinfo {author}
  {\bibfnamefont{A.}~\bibnamefont{Ishikawa}},\ and\ \bibinfo {author}
  {\bibfnamefont{Y.}~\bibnamefont{Arakawa}},\ }%
  \bibfield{journal}{%
  \bibinfo {journal} {Phys. Rev. Lett.}\ }%
  \textbf{\bibinfo {volume} {69}},\ \bibinfo {pages} {3314} (\bibinfo {year}
  {1992})%
  \bibAnnoteFile{NoStop}{Weisbuch1992}%
\bibitem{Baas2004}%
  \BibitemOpen
  \bibfield{author}{%
  \bibinfo {author} {\bibfnamefont{A.}~\bibnamefont{Baas}}, \bibinfo {author}
  {\bibfnamefont{J.~P.}\ \bibnamefont{Karr}}, \bibinfo {author}
  {\bibfnamefont{H.}~\bibnamefont{Eleuch}},\ and\ \bibinfo {author}
  {\bibfnamefont{E.}~\bibnamefont{Giacobino}},\ }%
  \bibfield{journal}{%
  \Doi{10.1103/PhysRevA.69.023809}{\bibinfo {journal} {Phys. Rev. A}}\ }%
  \textbf{\bibinfo {volume} {69}},\ \bibinfo {pages} {023809} (\bibinfo {month}
  {Feb}\ \bibinfo {year} {2004})%
  \bibAnnoteFile{NoStop}{Baas2004}%
\bibitem{Leyder2007}%
  \BibitemOpen
  \bibfield{author}{%
  \bibinfo {author} {\bibfnamefont{C.}~\bibnamefont{Leyder}}, \bibinfo {author}
  {\bibfnamefont{M.}~\bibnamefont{Romanelli}}, \bibinfo {author}
  {\bibfnamefont{J.~P.}\ \bibnamefont{Karr}}, \bibinfo {author}
  {\bibfnamefont{E.}~\bibnamefont{Giacobino}}, \bibinfo {author}
  {\bibfnamefont{T.}~\bibnamefont{Liew}}, \bibinfo {author}
  {\bibfnamefont{M.}~\bibnamefont{Glazov}}, \bibinfo {author}
  {\bibfnamefont{A.}~\bibnamefont{Kavokin}}, \bibinfo {author}
  {\bibfnamefont{G.}~\bibnamefont{Malpuech}},\ and\ \bibinfo {author}
  {\bibfnamefont{A.}~\bibnamefont{Bramati}},\ }%
  \bibfield{journal}{%
  \bibinfo {journal} {Nature Phys.}\ }%
  \textbf{\bibinfo {volume} {3}},\ \bibinfo {pages} {628} (\bibinfo {year}
  {2007})%
  \bibAnnoteFile{NoStop}{Leyder2007}%
\bibitem{Lagoudakis2009}%
  \BibitemOpen
  \bibfield{author}{%
  \bibinfo {author} {\bibfnamefont{K.~G.}\ \bibnamefont{Lagoudakis}}, \bibinfo
  {author} {\bibfnamefont{T.}~\bibnamefont{Ostatnicky}}, \bibinfo {author}
  {\bibfnamefont{A.~V.}\ \bibnamefont{Kavokin}}, \bibinfo {author}
  {\bibfnamefont{Y.~G.}\ \bibnamefont{Rubo}}, \bibinfo {author}
  {\bibfnamefont{R.}~\bibnamefont{Andr\'e}, \bibfnamefont{R.}},\ and\ \bibinfo
  {author} {\bibfnamefont{B.}~\bibnamefont{Deveaud-Pl\'edran}},\ }%
  \bibfield{journal}{%
  \Doi{10.1126/science.1177980}{\bibinfo {journal} {Science}}\ }%
  \textbf{\bibinfo {volume} {326}},\ \bibinfo {pages} {974} (\bibinfo {year}
  {2009})%
  \bibAnnoteFile{NoStop}{Lagoudakis2009}%
\bibitem{Hivet2012}%
  \BibitemOpen
  \bibfield{author}{%
  \bibinfo {author} {\bibfnamefont{R.}~\bibnamefont{Hivet}}, \bibinfo {author}
  {\bibfnamefont{H.}~\bibnamefont{Flayac}}, \bibinfo {author}
  {\bibfnamefont{D.}~\bibnamefont{Solnyshkov}}, \bibinfo {author}
  {\bibfnamefont{D.}~\bibnamefont{Tanese}}, \bibinfo {author}
  {\bibfnamefont{T.}~\bibnamefont{Boulier}}, \bibinfo {author}
  {\bibfnamefont{D.}~\bibnamefont{Andreoli}}, \bibinfo {author}
  {\bibfnamefont{E.}~\bibnamefont{Giacobino}}, \bibinfo {author}
  {\bibfnamefont{J.}~\bibnamefont{Bloch}}, \bibinfo {author}
  {\bibfnamefont{A.}~\bibnamefont{Bramati}}, \bibinfo {author}
  {\bibfnamefont{G.}~\bibnamefont{Malpuech}}, \emph{et~al.},\ }%
  \bibfield{journal}{%
  \bibinfo {journal} {Nature Physics}\ }%
  \textbf{\bibinfo {volume} {8}},\ \bibinfo {pages} {724} (\bibinfo {year}
  {2012})%
  \bibAnnoteFile{NoStop}{Hivet2012}%
\bibitem{Takemura2014}%
  \BibitemOpen
  \bibfield{author}{%
  \bibinfo {author} {\bibfnamefont{N.}~\bibnamefont{Takemura}}, \bibinfo
  {author} {\bibfnamefont{S.}~\bibnamefont{Trebaol}}, \bibinfo {author}
  {\bibfnamefont{M.}~\bibnamefont{Wouters}}, \bibinfo {author}
  {\bibfnamefont{M.~T.}\ \bibnamefont{Portella-Oberli}},\ and\ \bibinfo
  {author} {\bibfnamefont{B.}~\bibnamefont{Deveaud}},\ }%
  \bibfield{journal}{%
  \bibinfo {journal} {Nature Physics}\ }%
  \textbf{\bibinfo {volume} {10}},\ \bibinfo {pages} {500} (\bibinfo {year}
  {2014})%
  \bibAnnoteFile{NoStop}{Takemura2014}%
\bibitem{Gippius2007}%
  \BibitemOpen
  \bibfield{author}{%
  \bibinfo {author} {\bibfnamefont{N.~A.}\ \bibnamefont{Gippius}}, \bibinfo
  {author} {\bibfnamefont{I.~A.}\ \bibnamefont{Shelykh}}, \bibinfo {author}
  {\bibfnamefont{D.~D.}\ \bibnamefont{Solnyshkov}}, \bibinfo {author}
  {\bibfnamefont{S.~S.}\ \bibnamefont{Gavrilov}}, \bibinfo {author}
  {\bibfnamefont{Y.~G.}\ \bibnamefont{Rubo}}, \bibinfo {author}
  {\bibfnamefont{A.~V.}\ \bibnamefont{Kavokin}}, \bibinfo {author}
  {\bibfnamefont{S.~G.}\ \bibnamefont{Tikhodeev}},\ and\ \bibinfo {author}
  {\bibfnamefont{G.}~\bibnamefont{Malpuech}},\ }%
  \bibfield{journal}{%
  \Doi{10.1103/PhysRevLett.98.236401}{\bibinfo {journal} {Phys. Rev. Lett.}}\
  }%
  \textbf{\bibinfo {volume} {98}},\ \bibinfo {pages} {236401} (\bibinfo {month}
  {Jun}\ \bibinfo {year} {2007})%
  \bibAnnoteFile{NoStop}{Gippius2007}%
\bibitem{Paraiso2010}%
  \BibitemOpen
  \bibfield{author}{%
  \bibinfo {author} {\bibfnamefont{T.}~\bibnamefont{Paraiso}}, \bibinfo
  {author} {\bibfnamefont{M.}~\bibnamefont{Wouters}}, \bibinfo {author}
  {\bibfnamefont{Y.}~\bibnamefont{L{\'e}ger}}, \bibinfo {author}
  {\bibfnamefont{F.}~\bibnamefont{Morier-Genoud}},\ and\ \bibinfo {author}
  {\bibfnamefont{B.}~\bibnamefont{Deveaud-Pl{\'e}dran}},\ }%
  \bibfield{journal}{%
  \bibinfo {journal} {Nature mat.}\ }%
  \textbf{\bibinfo {volume} {9}},\ \bibinfo {pages} {655} (\bibinfo {year}
  {2010})%
  \bibAnnoteFile{NoStop}{Paraiso2010}%
\bibitem{Amo2010}%
  \BibitemOpen
  \bibfield{author}{%
  \bibinfo {author} {\bibfnamefont{A.}~\bibnamefont{Amo}}, \bibinfo {author}
  {\bibfnamefont{T.}~\bibnamefont{Liew}}, \bibinfo {author}
  {\bibfnamefont{C.}~\bibnamefont{Adrados}}, \bibinfo {author}
  {\bibfnamefont{R.}~\bibnamefont{Houdr{\'e}}}, \bibinfo {author}
  {\bibfnamefont{E.}~\bibnamefont{Giacobino}}, \bibinfo {author}
  {\bibfnamefont{A.}~\bibnamefont{Kavokin}},\ and\ \bibinfo {author}
  {\bibfnamefont{A.}~\bibnamefont{Bramati}},\ }%
  \bibfield{journal}{%
  \bibinfo {journal} {Nature Phot.}\ }%
  \textbf{\bibinfo {volume} {4}},\ \bibinfo {pages} {361} (\bibinfo {year}
  {2010})%
  \bibAnnoteFile{NoStop}{Amo2010}%
\bibitem{Cerna2013}%
  \BibitemOpen
  \bibfield{author}{%
  \bibinfo {author} {\bibfnamefont{R.}~\bibnamefont{Cerna}}, \bibinfo {author}
  {\bibfnamefont{Y.}~\bibnamefont{L{\'e}ger}}, \bibinfo {author}
  {\bibfnamefont{T.~K.}\ \bibnamefont{Para{\"\i}so}}, \bibinfo {author}
  {\bibfnamefont{M.}~\bibnamefont{Wouters}}, \bibinfo {author}
  {\bibfnamefont{F.}~\bibnamefont{Morier-Genoud}}, \bibinfo {author}
  {\bibfnamefont{M.~T.}\ \bibnamefont{Portella-Oberli}},\ and\ \bibinfo
  {author} {\bibfnamefont{B.}~\bibnamefont{Deveaud}},\ }%
  \bibfield{journal}{%
  \bibinfo {journal} {Nature Comm.}\ }%
  \textbf{\bibinfo {volume} {4}} (\bibinfo {year} {2013})%
  \bibAnnoteFile{NoStop}{Cerna2013}%
\bibitem{Tsintzos2008}%
  \BibitemOpen
  \bibfield{author}{%
  \bibinfo {author} {\bibfnamefont{S.}~\bibnamefont{Tsintzos}}, \bibinfo
  {author} {\bibfnamefont{N.}~\bibnamefont{Pelekanos}}, \bibinfo {author}
  {\bibfnamefont{G.}~\bibnamefont{Konstantinidis}}, \bibinfo {author}
  {\bibfnamefont{Z.}~\bibnamefont{Hatzopoulos}},\ and\ \bibinfo {author}
  {\bibfnamefont{P.}~\bibnamefont{Savvidis}},\ }%
  \bibfield{journal}{%
  \bibinfo {journal} {Nature}\ }%
  \textbf{\bibinfo {volume} {453}},\ \bibinfo {pages} {372} (\bibinfo {year}
  {2008})%
  \bibAnnoteFile{NoStop}{Tsintzos2008}%
\bibitem{Ballarini2013}%
  \BibitemOpen
  \bibfield{author}{%
  \bibinfo {author} {\bibfnamefont{D.}~\bibnamefont{Ballarini}}, \bibinfo
  {author} {\bibfnamefont{M.}~\bibnamefont{De~Giorgi}}, \bibinfo {author}
  {\bibfnamefont{E.}~\bibnamefont{Cancellieri}}, \bibinfo {author}
  {\bibfnamefont{R.}~\bibnamefont{Houdr\'e}}, \bibinfo {author}
  {\bibfnamefont{E.}~\bibnamefont{Giacobino}}, \bibinfo {author}
  {\bibfnamefont{R.}~\bibnamefont{Cingolani}}, \bibinfo {author}
  {\bibfnamefont{A.}~\bibnamefont{Bramati}}, \bibinfo {author}
  {\bibfnamefont{G.}~\bibnamefont{Gigli}},\ and\ \bibinfo {author}
  {\bibfnamefont{S.}~\bibnamefont{D.}},\ }%
  \bibfield{journal}{%
  \bibinfo {journal} {Nature Comm.}\ }%
  \textbf{\bibinfo {volume} {4}},\ \bibinfo {pages} {1778} (\bibinfo {year}
  {2013})%
  \bibAnnoteFile{NoStop}{Ballarini2013}%
\bibitem{Grosso2014}%
  \BibitemOpen
  \bibfield{author}{%
  \bibinfo {author} {\bibfnamefont{G.}~\bibnamefont{Grosso}}, \bibinfo {author}
  {\bibfnamefont{S.}~\bibnamefont{Trebaol}}, \bibinfo {author}
  {\bibfnamefont{M.}~\bibnamefont{Wouters}}, \bibinfo {author}
  {\bibfnamefont{F.}~\bibnamefont{Morier-Genoud}}, \bibinfo {author}
  {\bibfnamefont{M.}~\bibnamefont{Portella-Oberli}},\ and\ \bibinfo {author}
  {\bibfnamefont{B.}~\bibnamefont{Deveaud}},\ }%
  \bibfield{journal}{%
  \bibinfo {journal} {Physical Review B}\ }%
  \textbf{\bibinfo {volume} {90}},\ \bibinfo {pages} {045307} (\bibinfo {year}
  {2014})%
  \bibAnnoteFile{NoStop}{Grosso2014}%
\bibitem{Glazov2013}%
  \BibitemOpen
  \bibfield{author}{%
  \bibinfo {author} {\bibfnamefont{M.}~\bibnamefont{Glazov}}, \bibinfo {author}
  {\bibfnamefont{M.}~\bibnamefont{Semina}}, \bibinfo {author}
  {\bibfnamefont{E.~Y.}\ \bibnamefont{Sherman}},\ and\ \bibinfo {author}
  {\bibfnamefont{A.}~\bibnamefont{Kavokin}},\ }%
  \bibfield{journal}{%
  \bibinfo {journal} {Physical Review B}\ }%
  \textbf{\bibinfo {volume} {88}},\ \bibinfo {pages} {041309} (\bibinfo {year}
  {2013})%
  \bibAnnoteFile{NoStop}{Glazov2013}%
\bibitem{Poltavtsev2014}%
  \BibitemOpen
  \bibfield{author}{%
  \bibinfo {author} {\bibfnamefont{S.}~\bibnamefont{Poltavtsev}}, \bibinfo
  {author} {\bibfnamefont{I.}~\bibnamefont{Ryzhov}}, \bibinfo {author}
  {\bibfnamefont{M.}~\bibnamefont{Glazov}}, \bibinfo {author}
  {\bibfnamefont{G.}~\bibnamefont{Kozlov}}, \bibinfo {author}
  {\bibfnamefont{V.}~\bibnamefont{Zapasskii}}, \bibinfo {author}
  {\bibfnamefont{A.}~\bibnamefont{Kavokin}}, \bibinfo {author}
  {\bibfnamefont{P.}~\bibnamefont{Lagoudakis}}, \bibinfo {author}
  {\bibfnamefont{D.}~\bibnamefont{Smirnov}},\ and\ \bibinfo {author}
  {\bibfnamefont{E.}~\bibnamefont{Ivchenko}},\ }%
  \bibfield{journal}{%
  \bibinfo {journal} {Physical Review B}\ }%
  \textbf{\bibinfo {volume} {89}},\ \bibinfo {pages} {081304} (\bibinfo {year}
  {2014})%
  \bibAnnoteFile{NoStop}{Poltavtsev2014}%
\bibitem{Boulier2014}%
  \BibitemOpen
  \bibfield{author}{%
  \bibinfo {author} {\bibfnamefont{T.}~\bibnamefont{Boulier}}, \bibinfo
  {author} {\bibfnamefont{M.}~\bibnamefont{Bamba}}, \bibinfo {author}
  {\bibfnamefont{A.}~\bibnamefont{Amo}}, \bibinfo {author}
  {\bibfnamefont{C.}~\bibnamefont{Adrados}}, \bibinfo {author}
  {\bibfnamefont{A.}~\bibnamefont{Lemaitre}}, \bibinfo {author}
  {\bibfnamefont{E.}~\bibnamefont{Galopin}}, \bibinfo {author}
  {\bibfnamefont{I.}~\bibnamefont{Sagnes}}, \bibinfo {author}
  {\bibfnamefont{J.}~\bibnamefont{Bloch}}, \bibinfo {author}
  {\bibfnamefont{C.}~\bibnamefont{Ciuti}}, \bibinfo {author}
  {\bibfnamefont{E.}~\bibnamefont{Giacobino}}, \emph{et~al.},\ }%
  \bibfield{journal}{%
  \bibinfo {journal} {Nature Comm.}\ }%
  \textbf{\bibinfo {volume} {5}} (\bibinfo {year} {2014})%
  \bibAnnoteFile{NoStop}{Boulier2014}%
\bibitem{Murali2009}%
  \BibitemOpen
  \bibfield{author}{%
  \bibinfo {author} {\bibfnamefont{K.}~\bibnamefont{Murali}}, \bibinfo {author}
  {\bibfnamefont{S.}~\bibnamefont{Sinha}}, \bibinfo {author}
  {\bibfnamefont{W.~L.}\ \bibnamefont{Ditto}},\ and\ \bibinfo {author}
  {\bibfnamefont{A.~R.}\ \bibnamefont{Bulsara}},\ }%
  \bibfield{journal}{%
  \bibinfo {journal} {Physical Review Letters}\ }%
  \textbf{\bibinfo {volume} {102}},\ \bibinfo {pages} {104101} (\bibinfo {year}
  {2009})%
  \bibAnnoteFile{NoStop}{Murali2009}%
\bibitem{Liew2008}%
  \BibitemOpen
  \bibfield{author}{%
  \bibinfo {author} {\bibfnamefont{T.}~\bibnamefont{Liew}}, \bibinfo {author}
  {\bibfnamefont{A.}~\bibnamefont{Kavokin}},\ and\ \bibinfo {author}
  {\bibfnamefont{I.}~\bibnamefont{Shelykh}},\ }%
  \bibfield{journal}{%
  \bibinfo {journal} {Physical Review Letters}\ }%
  \textbf{\bibinfo {volume} {101}},\ \bibinfo {pages} {016402} (\bibinfo {year}
  {2008})%
  \bibAnnoteFile{NoStop}{Liew2008}%
\bibitem{Cheng2010}%
  \BibitemOpen
  \bibfield{author}{%
  \bibinfo {author} {\bibfnamefont{X.}~\bibnamefont{Cheng}}, \bibinfo {author}
  {\bibfnamefont{C.~T.}\ \bibnamefont{Boone}}, \bibinfo {author}
  {\bibfnamefont{J.}~\bibnamefont{Zhu}},\ and\ \bibinfo {author}
  {\bibfnamefont{I.~N.}\ \bibnamefont{Krivorotov}},\ }%
  \bibfield{journal}{%
  \bibinfo {journal} {Physical Review Letters}\ }%
  \textbf{\bibinfo {volume} {105}},\ \bibinfo {pages} {047202} (\bibinfo {year}
  {2010})%
  \bibAnnoteFile{NoStop}{Cheng2010}%
\bibitem{Finocchio2011}%
  \BibitemOpen
  \bibfield{author}{%
  \bibinfo {author} {\bibfnamefont{G.}~\bibnamefont{Finocchio}}, \bibinfo
  {author} {\bibfnamefont{I.}~\bibnamefont{Krivorotov}}, \bibinfo {author}
  {\bibfnamefont{X.}~\bibnamefont{Cheng}}, \bibinfo {author}
  {\bibfnamefont{L.}~\bibnamefont{Torres}},\ and\ \bibinfo {author}
  {\bibfnamefont{B.}~\bibnamefont{Azzerboni}},\ }%
  \bibfield{journal}{%
  \bibinfo {journal} {Physical Review B}\ }%
  \textbf{\bibinfo {volume} {83}},\ \bibinfo {pages} {134402} (\bibinfo {year}
  {2011})%
  \bibAnnoteFile{NoStop}{Finocchio2011}%
\bibitem{Gourier1998}%
  \BibitemOpen
  \bibfield{author}{%
  \bibinfo {author} {\bibfnamefont{D.}~\bibnamefont{Gourier}}\ and\ \bibinfo
  {author} {\bibfnamefont{D.}~\bibnamefont{Gerbault}},\ }%
  \bibfield{journal}{%
  \bibinfo {journal} {Physical Review B}\ }%
  \textbf{\bibinfo {volume} {57}},\ \bibinfo {pages} {2679} (\bibinfo {year}
  {1998})%
  \bibAnnoteFile{NoStop}{Gourier1998}%
\bibitem{Kramers1940}%
  \BibitemOpen
  \bibfield{author}{%
  \bibinfo {author} {\bibfnamefont{H.~A.}\ \bibnamefont{Kramers}},\ }%
  \bibfield{journal}{%
  \bibinfo {journal} {Physica}\ }%
  \textbf{\bibinfo {volume} {7}},\ \bibinfo {pages} {284} (\bibinfo {year}
  {1940})%
  \bibAnnoteFile{NoStop}{Kramers1940}%
\bibitem{ElDaif2006}%
  \BibitemOpen
  \bibfield{author}{%
  \bibinfo {author} {\bibfnamefont{O.}~\bibnamefont{El~Da{\"\i}f}}, \bibinfo
  {author} {\bibfnamefont{A.}~\bibnamefont{Baas}}, \bibinfo {author}
  {\bibfnamefont{T.}~\bibnamefont{Guillet}}, \bibinfo {author}
  {\bibfnamefont{J.-P.}\ \bibnamefont{Brantut}}, \bibinfo {author}
  {\bibfnamefont{R.~I.}\ \bibnamefont{Kaitouni}}, \bibinfo {author}
  {\bibfnamefont{J.-L.}\ \bibnamefont{Staehli}}, \bibinfo {author}
  {\bibfnamefont{F.}~\bibnamefont{Morier-Genoud}},\ and\ \bibinfo {author}
  {\bibfnamefont{B.}~\bibnamefont{Deveaud}},\ }%
  \bibfield{journal}{%
  \bibinfo {journal} {Applied Physics Letters}\ }%
  \textbf{\bibinfo {volume} {88}},\ \bibinfo {pages} {061105} (\bibinfo {year}
  {2006})%
  \bibAnnoteFile{NoStop}{ElDaif2006}%
\bibitem{Nardin2009}%
  \BibitemOpen
  \bibfield{author}{%
  \bibinfo {author} {\bibfnamefont{G.}~\bibnamefont{Nardin}}, \bibinfo {author}
  {\bibfnamefont{T.~K.}\ \bibnamefont{Para{\"\i}so}}, \bibinfo {author}
  {\bibfnamefont{R.}~\bibnamefont{Cerna}}, \bibinfo {author}
  {\bibfnamefont{B.}~\bibnamefont{Pietka}}, \bibinfo {author}
  {\bibfnamefont{Y.}~\bibnamefont{L{\'e}ger}}, \bibinfo {author}
  {\bibfnamefont{O.}~\bibnamefont{El~Daif}}, \bibinfo {author}
  {\bibfnamefont{F.}~\bibnamefont{Morier-Genoud}},\ and\ \bibinfo {author}
  {\bibfnamefont{B.}~\bibnamefont{Deveaud-Pl{\'e}dran}},\ }%
  \bibfield{journal}{%
  \bibinfo {journal} {Applied Physics Letters}\ }%
  \textbf{\bibinfo {volume} {94}},\ \bibinfo {pages} {181103} (\bibinfo {year}
  {2009})%
  \bibAnnoteFile{NoStop}{Nardin2009}%
\bibitem{Smith1977}%
  \BibitemOpen
  \bibfield{author}{%
  \bibinfo {author} {\bibfnamefont{P.}~\bibnamefont{Smith}}\ and\ \bibinfo
  {author} {\bibfnamefont{E.}~\bibnamefont{Turner}},\ }%
  \bibfield{journal}{%
  \bibinfo {journal} {Appl. Phys. Lett.}\ }%
  \textbf{\bibinfo {volume} {30}},\ \bibinfo {pages} {280} (\bibinfo {year}
  {1977})%
  \bibAnnoteFile{NoStop}{Smith1977}%
\bibitem{Wouters2013}%
  \BibitemOpen
  \bibfield{author}{%
  \bibinfo {author} {\bibfnamefont{M.}~\bibnamefont{Wouters}}, \bibinfo
  {author} {\bibfnamefont{T.~K.}\ \bibnamefont{Paraiso}}, \bibinfo {author}
  {\bibfnamefont{Y.}~\bibnamefont{L\'eger}}, \bibinfo {author}
  {\bibfnamefont{R.}~\bibnamefont{Cerna}}, \bibinfo {author}
  {\bibfnamefont{F.}~\bibnamefont{Morier-Genoud}}, \bibinfo {author}
  {\bibfnamefont{M.~T.}\ \bibnamefont{Portella-Oberli}},\ and\ \bibinfo
  {author} {\bibfnamefont{B.}~\bibnamefont{Deveaud-Pl\'edran}},\ }%
  \bibfield{journal}{%
  \Doi{10.1103/PhysRevB.87.045303}{\bibinfo {journal} {Phys. Rev. B}}\ }%
  \textbf{\bibinfo {volume} {87}},\ \bibinfo {pages} {045303} (\bibinfo {month}
  {Jan}\ \bibinfo {year} {2013})%
  \bibAnnoteFile{NoStop}{Wouters2013}%
\end{thebibliography}
%

\end{document}